\documentstyle[12pt]{article}

\newcommand{\eq}[1]{eq.(\ref{#1})}

\def\fm{~\mbox{fm}}
\def\tr{\mbox{Tr}}

\def\be{\begin{equation}}
\def\ee{\end{equation}}
\def\ltap{\raisebox{-.55ex}{\rlap{$\sim$}} \raisebox{.4ex}{$<$}}
\def\gtap{\raisebox{-.55ex}{\rlap{$\sim$}} \raisebox{.4ex}{$>$}}
\def\gsim{\mathrel{\gtap}}
\def\lsim{\mathrel{\ltap}}

\begin{document}
\begin{flushright}
UCLA/93/TEP/10\\
April 1993\\
hep-ph/9305207
\end{flushright}
\vspace{0.5in}
\begin{center}
{\Large Disoriented chiral condensate in (1+1) Lorentz-invariant
geometry}\\
\vspace{0.4in}
{\large S.Yu.~Khlebnikov}
\footnote{On leave of absence
from Institute for Nuclear Research of the Academy
of Sciences, Moscow 117312 Russia.} \\
\vspace{0.2in}
{\it Department of Physics, University of California,
Los Angeles, CA 90024, USA} \\
\vspace{0.7in}
{\bf Abstract} \\
\end{center}
We consider isospin correlations of pions produced in a relativistic
nuclear collision, using an effective theory of the chiral order
parameter. Our theory has (1+1) Lorentz invariance as appropriate
for the central rapidity region. We argue that in certain regions of space
correlations of the chiral order parameter
are described by the fixed point of the (1+1) WZNW model.
The corresponding anomalous dimension determines scaling of the
probability to observe a correlated cluster of pions
with the size of this cluster in rapidity.
Though the maximal size of clusters for which this scaling is
applicable is cut off by pion mass, such clusters can still include
sufficiently many particles to make the scaling observable.

%\vspace{0.3in} \\
%\noindent
\newpage
The Centauro events observed in cosmic ray experiments
\cite{experiment} stimulated interest in unusual states of hadronic
matter \cite{explain}. A now popular idea is that of disoriented
chiral condensate \cite{ARBK,proposal,RW}: one assumes that in course of a
relativistic nuclear collision, relatively large domains of nuclear matter
may emit pions coherently. An experimental signature of such correlated
domains would be that certain regions in the angle-rapidity plane are
overpopulated by pions of the same isospin, say, there are
regions containing mostly $\pi^{\pm}$'s without many of
$\pi^{0}$'s, or vice versa. There is a proposal to look for such
events at the SSC \cite{proposal}.
These ideas are now discussed not only in connection with leading-particle
effects but also for the central rapidity region.
Rajagopal and Wilczek \cite{RW} consider formation of correlated domains
at the second-order QCD phase transition. They argue that
unless temperature drops very fast, the second-order phase transition
is unlikely to lead to large correlated domains because
the growing correlation length is cutoff by the pion mass. On the
other hand, for a sudden quench from high to low temperatures, they
find evidence for amplification of long-wavelength pion modes.

The results of ref.\cite{RW} are obtained within a (3+1)-dimensional
effective theory. One may wonder whether
this is always an adequate dimensionality. Indeed,
the analysis of experimental data \cite{Bjorken,KM}, in particular
those from the CERN SPS collider \cite{UA5}, indicates that the
distribution of particles in the central rapidity region possesses (1+1)
Lorentz invariance with respect to the longitudinal coordinate $z$ and
time $t$. This means that the distribution looks alike
in all reference frames obtained from the center-of-mass (c.m.) frame by
modest Lorentz boosts in the longitudinal direction.
On the other hand, the nuclear matter produced in a collision does not
maintain either (3+1)
Lorentz invariance or the three-dimensional rotation symmetry.
We will argue that in these circumstances,
the most long-range part of isospin correlations is described by a
(1+1)-dimensional effective theory. We also discuss a cutoff
role of the pion mass in this case.
Our analysis is simpler for the central rapidity region and
we concentrate on this region in what follows. This does not exclude
a possibility of similar correlation phenomena at larger rapidities.

We need to know statistical weights of various chiral configurations
in the regions of physical space producing the central rapidity particles.
To maintain the (1+1) Lorentz symmetry, these statistical weights
should be determined by a (1+1) Lorentz-invariant effective action.
For low-momentum low-frequency modes, the essentially unique choice
is a variant of the non-linear sigma model.\footnote{Unlike ref.\cite{RW},
we do not use a linear sigma model because we expect our system to
be disordered by fluctuations of the phases of the order parameter,
rather than of its magnitude.} Since we do not have
to maintain the full (3+1) Lorentz invariance in this effective
theory, more terms
become allowed in the chiral lagrangian. In particular, now there are two
terms of the second order in derivatives with respect to $z$ and $t$.
They form the action
\be
S =\frac{f^2}{4} \int d^4 x ~\tr \partial_{\mu} u
\partial_{\mu} u^{-1} + g^2 \int d^2 x_{\perp} \Gamma_2 \; ,
\label{action}
\ee
where $\mu,\nu=z,t$,
$u=\exp(i\sum\tau^a \pi^a/f)$, $\pi^a$ are the three pion fields,
$\tau^a$ are the Pauli matrices and $x_{\perp}$ are coordinates in the
transverse plane. We will discuss the region of transverse integration below.
The quantity $\Gamma_2$ is the two-dimensional Wess-Zumino term
\cite{Novikov,Witten}; it is
compatible with both Lorentz and chiral symmetries. For small $\pi^a$,
the Wess-Zumino term has the form
\be
\Gamma_2 \approx \frac{1}{6\pi f^3}
\int dz dt \epsilon^{\mu\nu} \epsilon^{abc}
\pi^a\partial_{\mu} \pi^b \partial_{\nu} \pi^c \; .
\label{gamma2}
\ee
In addition to (\ref{action}),
there are terms containing derivatives with respect to $x_{\perp}$;
the one lowest in derivatives is
\be
\frac{1}{4}\int d^4 x K_{\alpha\beta}
{}~\tr \partial_{\alpha} u \partial_{\beta} u^{-1} \; ,
\label{add}
\ee
where $\alpha,\beta=x_{\perp 1},x_{\perp 2}$
and $K_{\alpha\beta}$ is some $2\times 2$ tensor.
Note that due to the (1+1) Lorentz invariance, there are no second-order
in derivatives terms that would mix longitudinal and transverse indices.

Because we want to study multiparticle, and in this sense long-range,
correlations, we are interested in the values of the couplings
$f$, $g$ and $K_{\alpha\beta}$ attained after an appropriate renormalization.
A suitable renormalization procedure in our case is the scaling
$z\to \lambda z$ and $t\to \lambda t$ leaving the transverse coordinates
$x_{\perp}$ unchanged. Note that the longitudinal and transverse couplings
can renormalize differently.
By simple counting of dimensions, the transverse coupling $K_{\alpha\beta}$
is important when $K_{\alpha\beta} \gsim \mu^2$, where $\mu$ is a
normalization mass. At $K_{\alpha\beta} \ll \mu^2$ the scaling laws
of the longitudinal theory (\ref{action}) that describes the
(1+1)-dimensional fluctuations renormalizing the couplings,
are those of the (1+1) WZNW model (see below). The chiral field $u$ scales
as $u\propto \mu^{-\gamma/2}$; for $\gamma$ we take the level $k=1$
value $\gamma=1$. The effective coupling
$K(\mu)$ then scales as $\mu^{\gamma}$ and therefore decreases in infrared.
Since $K$ decreases slower than $\mu^2$, at sufficiently small $\mu$
the dimensionless ratio $K/\mu^2$ becomes large and renormalization becomes
complicated. In the real world, however, the accessible values of $\mu$ are
bounded from below by the mass of pion. We will see that the
renormalized value of pion mass to be used in our case is
$m\approx 73~\mbox{MeV}$. The renormalized value of $K$ at $\mu=m$ is
\be
K_{\alpha\beta} (m) = K_{\alpha\beta} (\Lambda)
\left( \frac{m}{\Lambda} \right)^{\gamma} =
(26~\mbox{MeV})^2 \delta_{\alpha\beta} \; ,
\label{K}
\ee
where we assume that the cutoff is $\Lambda= 1~\mbox{GeV}$, and
$K_{\alpha\beta} (\Lambda)=f_{\pi}^2 \delta_{\alpha\beta}$,
$f_{\pi}=95~\mbox{MeV}$. Since $K(m)\ll m^2$, we may approximate the
behavior of the system at longitudinal distances smaller than $m^{-1}$ by
the theory at the fixed point $K_{\alpha\beta}=0$.

At $K_{\alpha\beta}=0$ fluctuations of the order
parameter in different regions of the transverse plane decouple
from each other and form infinitely many one-dimensional systems.
It turns out that long-range correlations in rapidity develop only
in some regions of the transverse plane and only for
a particular transverse size. We call this favorable size $d$ and will
estimate it shortly to be $d\approx 0.8\fm$.
Correlations for other transverse sizes are short-ranged.
The size $d$ has the significance of an infrared fixed point:
coarse graining in proper time aligns the order parameter
in certain transverse regions over the transverse size $d$.
Integrating \eq{action} over a transverse region of this
size gives a (1+1)-dimensional theory
\be
S =\frac{1}{2\lambda^2} \int dz dt ~\tr \partial_{\mu} u
\partial_{\mu} u^{-1} + k \Gamma_2 \; .
\label{2d}
\ee
This is the WZNW model \cite{Novikov,Witten}.
Here $\lambda=\sqrt{2}/(fd)$ is a dimensionless
coupling and $k$ is an apriori quantized, integer coefficient.
Let us temporarily neglect the mass effects. The theory (\ref{2d})
has an infrared stable fixed point at
$\lambda^2=4\pi/|k|$ \cite{Witten,PW}. This gives a value of the
transverse scale $d$ for which the rapidity correlations are most
long-ranged,
\be
d= \left( \frac{|k|}{2\pi} \right)^{1/2} \frac{1}{f} \; .
\label{d}
\ee
There may be some dependence of $f$ and $d$ on the energy but it need not
be dramatic. For estimates, we take $f = f_{\pi} = 95~\mbox{MeV}$.
We now have to determine the integer $k$.

The actual presence of the Wess-Zumino term in the effective theory
(\ref{action}), though allowed by the symmetries we discussed
so far, is not at all obvious.
This term breaks $G$-parity of pions, so it cannot happen in
a system consisting of pions alone. In the real system, however,
there are other degrees of freedom, and $G$-parity can be broken in
interactions of pions with them.
For quantized coefficients like $k$, the effects of heavier particles
are not necessarily suppressed by inverse powers of their masses.
The next lowest degree of freedom in nuclear matter are kaons.
In the (3+1) dimensional $SU(3)$ (that is including kaons) chiral
lagrangian, $G$-parity of pions is broken by the (3+1) Wess-Zumino term
\cite{WZW}. For small fields, the (3+1) Wess-Zumino action has the form
\be
N_c \Gamma_4 \approx \frac{N_c}{240\pi^2}
\int d^4 x \epsilon^{\mu\nu\rho\sigma} \tr
\Pi\partial_{\mu} \Pi\partial_{\nu} \Pi\partial_{\rho}
\Pi\partial_{\sigma}\Pi \; ,
\label{4d}
\ee
where $N_c=3$ is the number of colors, and the $3\times 3$ matrix $\Pi$
is a parametrization of an $SU(3)$ matrix $U$, $U=\exp(i\Pi)$.
We may write $\Pi=a+A$ where $a$ includes only pions and $A$ includes
only kaons and $\eta$.
Eq.(\ref{4d}) plays the role of a microscopic action with respect to
our effective theory. We have to make a reduction of \eq{4d}
to the fewer fields and fewer dimensions. This means that we need
a portion of \eq{4d} where $z$ and $t$ derivatives are applied only to
pions but not to kaons or $\eta$.
When the transverse integration is restricted to a correlated region of
size $d$, the required portion of $\Gamma_4$ factorizes,
\be
\Gamma_4\to -\frac{1}{4} \Gamma_2  Q \; ,
\label{factor}
\ee
where $\Gamma_2$ is the (1+1) Wess-Zumino term (\ref{gamma2}) and
\be
Q\approx -\frac{i}{2\pi} \int d^2 x_{\perp} \epsilon^{\alpha\beta}\tr
P\partial_{\alpha} A \partial_{\beta} A \; ,
\label{Q}
\ee
$P$ being a projector $3\times 3$ matrix, $P=\mbox{diag} (1,1,0)$.
The $\eta$ field does not contribute to \eq{Q}, so the matrix $A$
in this equation may be regarded as made purely of kaons.
Kaons span the projective space $SU(3)/(SU(2)\times U(1))\approx CP^2$,
and $Q$ is the integral over the correlated region
of the topological density of mappings from
the transverse plane into $CP^2$ \cite{CP2}.
Eq.(\ref{Q}) is its small-field limit. Therefore, the (3+1)
Wess-Zumino action produces a non-zero (1+1) Wess-Zumino action only
in the regions where
the transverse dependence of kaon fields is topologically non-trivial.
These topologically non-trivial configurations of kaons need
not be stable; we assume that they can live sufficiently long
to give rise to observable effects. Since the integration in \eq{Q}
is over a finite transverse region, $Q$ need not be integer. The minimal
value of $Q$ for which our dimensional reduction is consistent is
$Q=\pm 4/3$, corresponding to $k=\mp 1$ in \eq{2d}.
Then, for the transverse scale $d$, \eq{d} gives
$d\approx 0.8\fm$. The inverse of this determines a typical transverse
momentum of pions produced in correlated regions,
$d^{-1}\sim 240 ~\mbox{MeV}$.

At the critical point, that is for the favorable transverse size,
the longitudinal correlations in the theory (\ref{2d}) are scale-invariant.
The (unnormalized) probability $P(z,z';t,t')$ to have the order parameter
$u$ aligned over a distance from $z$ to $z'$ for a time from $t$ to
$t'$ is proportional to a two-point correlation function,
\be
P(z,z';t,t')~\propto ~\langle u(z,t) ~u^{-1}(z',t') \rangle ~\propto
{}~[(z-z')^2 - (t-t')^2]^{-\gamma/2} \; .
\label{P}
\ee
The anomalous dimension $\gamma$ is known \cite{KZ},
$\gamma=3/(k+2)$; for $k=1$, $\gamma=1$. We are interested in
the probability (\ref{P}) at the value of proper time $\tau=\tau_1$
at which the one-dimensional expansion changes into three-dimensional
and the system rapidly hadronizes. This time is proportional to nuclear
radius; for nuclei of Pb or U, Bjorken \cite{Bjorken} estimates it to be
$\tau_1\sim 7 \fm/c$.
Going over to the proper time $\tau$ and rapidity $y$ via
$z=\tau\sinh y$ and
$t=\tau\cosh y$, setting $\tau=\tau'=\tau_1$, and assuming $y-y'\ll 1$,
we get
\be
P(y,y')\propto \frac{1}{|y-y'|^{\gamma}} \; .
\label{Pw}
\ee
Eq.(\ref{Pw}) gives a distribution in rapidity sizes for clusters of
pions with a given neutral-to-charged ratio, as long as the mass effects
are neglected.

In actual comparison of \eq{Pw} with experimental data,
to separate the correlation predicted by (\ref{Pw}) from statistical
fluctuations, one has to select high-multiplicity events that have
clusters of pions with a far from average neutral-to-charged ratio.
For example, one selects clusters with more than 90 or less than 5 percent
of $\pi^0$'s. Then, one makes a histogram of occurrences of such clusters
in all of the selected events
versus their sizes in rapidity. As will be seen below, for very large
clusters, the probability is cut off by mass effects. However, for clusters
whose extent in rapidity is smaller than a certain $\Delta y$, the histogram
should show the scaling behavior of probability (\ref{Pw}) with the
universal exponent $\gamma$.

Now let us consider the role of pion mass and the possibility
to observe correlations \eq{Pw} in the real world. The mass
cutoff factor now appearing in \eq{Pw} is
\be
\exp\left(-m [(z-z')^2 - (t-t')^2]^{1/2} \right) \approx
\exp(-m\tau_1 |y-y'|) \; .
\label{cutoff}
\ee
The renormalized value of pion mass is determined from the scaling equation
\be
m^2(m) = m^2(\Lambda) \left( \frac{m(m)}{\Lambda} \right)^{\gamma/2} \; .
\label{mass}
\ee
For $\Lambda=1~\mbox{GeV}$, $m(\Lambda)=140~\mbox{MeV}$,
$\gamma=1$, we find $m=73~\mbox{MeV}$.

Eq.(\ref{cutoff}) shows that the longer is
the one-dimensional evolution of the system, the smaller is the maximal
size of clusters for which the scaling law (\ref{Pw}) can be observed,
\be
\Delta y \lsim (m\tau_1)^{-1} \;
\label{interval}
\ee
(assuming $\Delta y \lsim 1$).
This conclusion itself is not specific for our particular (1+1) dimensional
theory; it holds whenever the (1+1) Lorentz invariance is preserved.
It is also not unsimilar to the conclusion reached by Rajagopal and Wilczek
\cite{RW} in the three-dimensional case if we identify the time $\tau_1$
as the "cooling" time of the system. There is, however, an important
distinction between the thermalized system of ref.\cite{RW} and the
present case. If the system is thermalized, the number of particles
produced by the correlation volume is essentially independent of the
energy of the initial colliding particles. On the other hand, the number
of particles in a fixed rapidity interval is known experimentally to
grow with energy. Selecting high-energy events with large multiplicities
per unit rapidity, one may be able to observe correlations (\ref{Pw})
even if the interval (\ref{interval}) is relatively small (see below).
Increasing the atomic number of colliding nuclei, on the other hand,
does not necessarily enhance correlations; in fact, it may even suppress
them. Indeed, in our theory, a correlated domain is
characterized by the fixed transverse size $d\sim 0.8\fm$. According to
estimates based on the hydrodynamic approach \cite{Bjorken}, multiplicity
per unit rapidity grows with the atomic number $A$ proportionally to the
total transverse area, $dN/dy\propto A^{2/3}$. Hence, the portion of
$dN/dy$ produced by the {\em correlated} transverse region is independent
of $A$. At the same time, $\tau_1\propto A^{1/3}$, so the number
of correlated particles in the rapidity interval (\ref{interval})
decreases as $A^{-1/3}$. In these circumstances, the best place to observe
the correlations (\ref{Pw}) is high-energy nucleon-nucleon
collisions (SSC), rather
than nuclei-nuclei collisions. One should keep in mind, however,
that using different estimates for the dependence of $dN/dy$ on $A$ existing
in the literature \cite{ARBK} may change this conclusion. For hadron-hadron
collisions, $\tau_1\sim 1\fm/c$, so the correlations extend over the
rapidity interval $\Delta y\sim 2.2$. It seems sufficient to have 15-20
particles per unit rapidity to be able to observe these correlations and
deduce the scaling exponent.

To summarize, we suggest that hadronic matter
created in relativistic nuclear collisions is best described not
by an effective thermal theory but by an effective theory possessing
(1+1) Lorentz invariance. The basis for this suggestion is the observed
Lorentz symmetry of the distribution of particles in the central rapidity
region. We argue that in certain regions of space
correlations of the chiral order parameter
are described by the fixed point of the WZNW model.
The corresponding anomalous dimension determines the scaling of the
probability to observe an isospin-correlated  cluster of pions
with the size of this cluster in rapidity.
Though the maximal size of clusters for which this scaling is
applicable is cut off by pion mass, such clusters can still include
sufficiently many particles to make the scaling observable.

I am grateful to R. D. Peccei for getting me interested in the subject
and for discussions and criticism,
to S. Chakravarty for discussions on critical fluctuations
and renormalization group, and to K. Aoki, P. van Driel, B. Gradwohl and
M. von Ins for discussions on various other topics related to this paper.
The author is supported by the Julian Schwinger fellowship at UCLA.


\begin{thebibliography}{99}
\bibitem{experiment}
C. M. G. Lattes, Y. Fujimoto and S. Hasegawa, Phys. Rep. {\bf 65},
151 (1980).
\bibitem{explain}
J. Bjorken and L. McLerran, Phys. Rev. {\bf D20}, 2353 (1979).
\bibitem{ARBK}
A. A. Anselm and M. G. Ryskin, Phys. Lett. {\bf B266}, 482 (1991);
J.-P. Blaizot and A. Krzywicki, Phys. Rev. {\bf D46}, 246 (1992).
\bibitem{proposal}
J. D. Bjorken, Int. J. Mod. Phys. {\bf A7}, 4189 (1992);
Acta Physica Polonica {\bf B23}, 561 (1992);
K. L. Kowalski and C. C. Taylor, Case Western Reserve University preprint
92-6, hep-ph/9211282 (1992).
\bibitem{RW}
K. Rajagopal and F. Wilczek, Princeton preprints PUPT-1347,
IASSNS-HEP-92/60, hep-ph/9210253 (1992) and PUPT-1389, IASSNS-HEP-93/16,
hep-ph/9303281 (1993).
\bibitem{Bjorken}
J. D. Bjorken, Phys. Rev. {\bf D27}, 140 (1983).
\bibitem{KM}
K. Kajantie and L. McLerran, Nucl. Phys. {\bf B214}, 261 (1983).
\bibitem{UA5} UA5 Collaboration, Phys. Rep. {\bf 154}, 247 (1987).
\bibitem{Novikov}
S. P. Novikov, Usp. Mat. Nauk, {\bf 37}, 3 (1982).
\bibitem{Witten}
E. Witten, Comm. Math. Phys. {\bf 92}, 455 (1984).
\bibitem{PW}
A. M. Polyakov and P. B. Wiegmann, Phys. Lett. {\bf 141B}, 223 (1984).
\bibitem{WZW}
J. Wess and B. Zumino, Phys. Lett. {\bf 37B}, 95 (1971);
E. Witten, Nucl. Phys. {\bf B223}, 422 (1983).
\bibitem{CP2}
A. D'Adda, M. L\"{u}scher and P. Di Vecchia, Nucl. Phys. {\bf B146}, 63
(1978); V. Golo and A. M. Perelomov, Phys. Lett. {\bf 79B}, 112
(1978); A. J. MacFarlane, Phys. Lett. {\bf 82B}, 239 (1979).
\bibitem{KZ}
V. G. Knizhnik and A. B. Zamolodchikov, Nucl. Phys. {\bf B247}, 83
(1984).

\end{thebibliography}
\end{document}